\documentclass[twocolumn,showpacs,preprintnumbers,amsmath,amssymb]{revtex4-1}
\usepackage{graphicx}
\usepackage{textcomp}

\begin{document}
\title{Coherent perfect absorption in a homogeneously-broadened two-level medium}
  \normalsize
\author{Stefano Longhi 
}
\address{Dipartimento di Fisica, Politecnico di Milano, Piazza L. da Vinci
32, I-20133 Milano, Italy}

%
\bigskip
\begin{abstract}
\noindent In recent works, it has been shown rather generally that
the time-reversed process of lasing at threshold realizes a coherent
perfect absorber (CPA). In a CPA, a lossy medium in an optical
cavity with a specific degree of dissipation, equal in modulus to
the gain of the lasing medium,  can perfectly absorb coherent
optical waves that are the time-reversed counterpart of the lasing
field. Here the time-reversed process of lasing is considered in
details for a homogeneously-broadened two-level medium in an optical
cavity, and the conditions for CPA are derived. It is shown that,
owing to the dispersive properties of the two-level medium, exact
time-reversal symmetry is broken and the frequency of the field at
which CPA occurs is generally different than the one of the lasing
mode. Moreover,  at a large cooperation parameter the observation of
CPA in the presence of bistability requires to operate in the upper
branch of the hysteresis cycle.
\end{abstract}

\pacs{42.55.Ah, 42.65.Pc, 42.65.Sf}


\maketitle

{\it Introduction.} It is well-known that the absorption properties
of a lossy medium can be conveniently controlled by optical
feedback. Several early studies have investigated coherent
absorption effects in optical cavities \cite{A1,A2,A3,A4}, with
applications to the realization of special modulators and
cavity-enhanced detectors \cite{A1,A2,A3}. The transmission
properties of an optical cavity filled by an absorbing medium have
also attracted great attention since long time in connection to
optical bistability (see, for instance, \cite{B1,B2,B3} and
references therein). In recent works, the idea of controlling the
properties of absorption of a lossy medium in presence of feedback
has been renewed \cite{StonePRL2010,StoneScience2011}, and the
unusual scattering properties of gain/absorber systems with certain
symmetries have been highlighted \cite{LonghiPRA2010,StonePRL2011}.
In particular, in Ref. \cite{StonePRL2010} it was proven rather
generally that a coherent perfect absorber (CPA) realizes the
time-reversed process of lasing at threshold \cite{StonePRL2010},
whereas in Ref. \cite{StoneScience2011} an experimental
demonstration of interferometric control of the absorption was
reported using a thin slice of silicon illuminated by two beams
\cite{StoneScience2011}. The main idea underlying CPA is that, since
in a steady-state process time reversal corresponds to interchanging
incoming and outgoing fields, the time-reversed process of  lasing
at threshold corresponds to perfect absorption of certain incoming
coherent light fields. In the reverse process of lasing at
threshold, the gain medium in the resonator is replaced by a lossy
medium, corresponding to a positive imaginary refractive index equal
in absolute value to that at the lasing threshold. Then, whenever
the system is illuminated coherently and monochromatically by the
time-reverse of the output of a lasing mode, the incident radiation
is perfectly absorbed \cite{StonePRL2010}. Such a simple picture of
CPA strictly holds provided that phenomena like saturation of the
absorption and absorption-induced dispersion in the medium
(responsible to the well-known frequency pulling effect in the
time-reversed process of lasing) are negligible. While the former
condition can be satisfied for low-power fields, the latter
phenomenon (being a linear one) occurs also for low input powers and
can be non-negligible when absorption exploits a
 narrow resonance frequency of the atomic medium. In this Report we consider the time-reversed process of lasing
  for a homogeneously-broadened two-level medium
in an optical cavity, and discuss in details the conditions for CPA
when saturation of absorption and loss-induced dispersion are
properly considered in the model. Owing to the dispersive properties
of the two-level medium, it is shown that the frequency of the field
at which CPA occurs is generally different than the one of the
lasing mode, and that a somewhat singular case appears when the
cavity photon and polarization decay rates are equal. Moreover, it
is shown that, for a strong cooperation parameter leading to optical
bistability, the observation of CPA requires to operate in the upper
branch of the bistable cycle.\\

\par
{\it The model.} Let us consider a standard model describing light
absorption/amplification in a homogeneously-broadened two-level
medium embedded in an optical cavity, similar to the one encountered
in the theory of optical bistability \cite{B3,B4} or in the
semiclassical theory of homogeneously-broadened lasers with an
injected signal \cite{L1}. Specifically, we consider a two-level
medium with a resonance frequency $\omega_0$ of length $l$ placed in
a ring cavity of total length $\mathfrak{L}$ which is coupled to the
outside by a single
 mirror of power transmittance $T$ [see Fig.1(a)]. An input
field of frequency $\omega$, close to $\omega_0$, and amplitude
$\mathcal{E}_0$ can be injected into the cavity, as shown in
Fig.1(a). As compared to the most common configurations considered
in the theory of optical bistability \cite{B3,B4}, the optical
cavity considered here is coupled to the outside by a single coupler
[mirror 1 in Fig.1(a)], whereas all other mirrors are assumed to
have $100 \%$ reflectivity. For such a cavity, the lasing and its
time-reversed counterpart, i.e. CPA, are schematically shown in
Figs.1(b) and (c). To study the operation of laser and CPA, let us
indicate by $\mathcal{E}(z,t)=(1/2) [ (\sqrt{\gamma_{\parallel}
\gamma_{\perp}} \hbar/ \mu ) A(z,t) \exp(i \omega t-ikz) +c.c.]$,
$\mathcal{\mathcal{P}}(z,t)=(1/2) [ (i \mu N_e
\sqrt{\gamma_{\parallel}/ \gamma_{\perp}}) \Lambda (z,t) \exp(i
\omega t-ikz) +c.c.]$ and $N(z,t)=N_e n(z,t)$ the electric field,
macroscopic polarization, and population difference in the medium,
respectively, where $z$ is the longitudinal spatial coordinate along
the ring, $\gamma_{\parallel}$ and $\gamma_{\perp}$ are the
population and dipole decay rates, respectively, $\mu$ is the
modulus of the electric dipole moment of the atoms, and $N_e$ is the
population difference at the equilibrium between the lower and upper
atomic levels. For an absorber, $N_e$ is positive and equal to the
total atomic population $N_t$, whereas for an amplifying medium
$N_e$ is negative and its value is determined by the pumping rate.
The evolution of the slowly-varying amplitudes $A$ and $\Lambda$ of
electric field and polarization in the medium are governed by the
Maxwell-Bloch equations \cite{B3,B4,L1}
\begin{eqnarray}
\partial_t A & = & -c \partial_z A + c \alpha \Lambda \\
\partial_t \Lambda & = & -\gamma_{\perp} \left[ (1+i\Delta)\Lambda+nA
\right] \\
\partial_t n & = & -\gamma_{\parallel} \left[ n-1-\frac{1}{2}(A^*\Lambda+A
\Lambda^*)\right] \; ,
\end{eqnarray}
\begin{figure}
\includegraphics[scale=0.53]{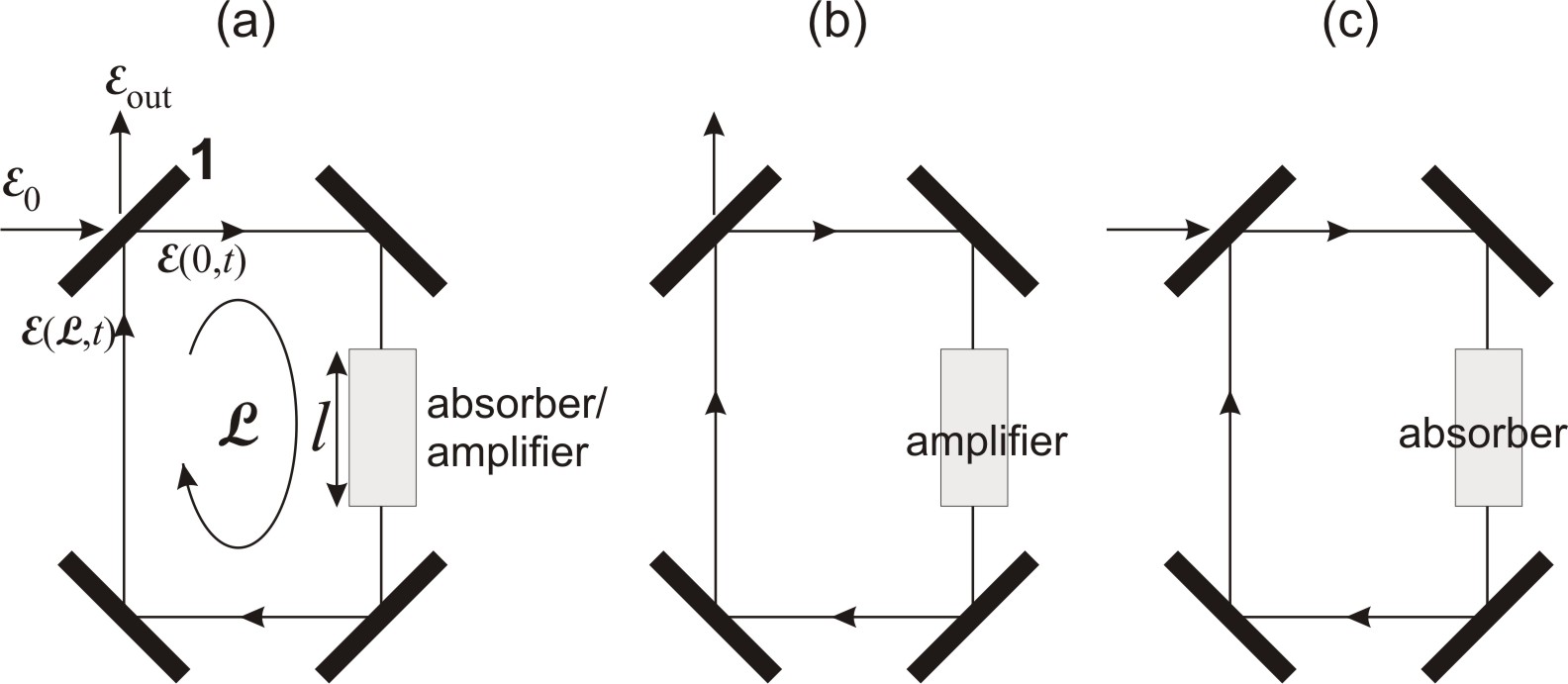}
\caption{(Color online) (a) Schematic of a ring cavity with a
homogeneously-broadened two-level medium and an injected signal. (b)
and (c) show schematically the operation of a laser and a CPA
system.}
\end{figure}
where $\alpha \equiv \mu^2\omega_0 N_e/(2 \hbar \gamma_{\perp}
\epsilon_0 c)$ is the unsaturated absorption ($N_e>0$) or
amplification ($N_e<0$) coefficient and $\Delta \equiv
(\omega-\omega_0) / \gamma_{\perp}$ is the detuning between the
frequency $\omega$ of the injected field and the resonance frequency
$\omega_0$ of the atoms, normalized to the polarization decay rate
$\gamma_{\perp}$ (i.e. to the gain/absorption linewidth). Equations
(1-3) should be supplemented by the boundary conditions imposed by
the coupling mirror 1 at $z=0$, which relates the incident and
scattered fields [see Fig.1(a)]. For a lossless mirror, these are
given by
\begin{equation}
A(0,t)=i \sqrt{T} A_0+\sqrt{1-T}A(\mathcal{L},t)
\exp(-ik\mathcal{\mathcal{L}})
\end{equation}
and
\begin{equation}
A_{out}=\sqrt{1-T} A_0+i \sqrt{T}A(\mathcal{L},t)
\exp(-ik\mathcal{\mathcal{L}})
\end{equation}
where $A_0 \equiv \mu \mathcal{E}_0 /(\hbar \sqrt{\gamma_{\parallel}
\gamma_{\perp}})$ and $A_{out} \equiv \mu \mathcal{E}_{out} /(\hbar
\sqrt{\gamma_{\parallel} \gamma_{\perp}})$ are the normalized
amplitudes of the injected and reflected electric fields at mirror
1, respectively [see Fig.1(a)]. Note that the laser ($\alpha<0$) and
CPA ($\alpha>0$) systems of Figs.1(b) and (c) correspond to
$\mathcal{E}_0=0, \; \mathcal{E}_{out} \neq 0$ and $\mathcal{E}_0
\neq 0, \; \mathcal{E}_{out}= 0$, respectively. To simplify the
analysis, we consider here the mean-field limit of the Maxwell-Bloch
equations \cite{B3,B4,L1}, in which the fields are almost uniform
along $z$. This limit requires a small transmissivity $T \rightarrow
0$, a small single-pass absorption/amplification coefficient $\alpha
l \rightarrow 0$ (of order $\sim T$), a small amplitude of the
injected signal $A_0 \rightarrow 0$ (of order $\sim \sqrt T$) and a
small detuning $k\mathcal{L}-2 n \pi \rightarrow 0$ (of order $\sim
T$), where $n$ is an integer defining the resonance frequency of the
cold cavity closest to $\omega$. In this limit, the Maxwell-Bloch
equations read
\begin{eqnarray}
\partial_t A & = & \kappa \left[ 2C \Lambda-(1+i\theta)A+Y \right] \\
\partial_t \Lambda & = & -\gamma_{\perp} \left[ (1+i\Delta) \Lambda+nA \right] \\
\partial_t n & = & -\gamma_{\parallel} \left[ n-1-\frac{1}{2}(A^*\Lambda+A
\Lambda^*)\right] \; ,
\end{eqnarray}
where $Y \equiv 2 A_0 i / \sqrt T$, $C \equiv \alpha l/T$ is the
cooperation parameter, $\kappa \equiv cT/(2\mathcal{L})$ is the
photon decay rate in the cold cavity, and $\theta \equiv
(\omega-\omega_c)/ \kappa$ is the detuning between the frequency
$\omega$ of the injected field and the cavity resonance frequency
$\omega_c=2 n \pi c/\mathcal{L}$ closest to $\omega$, normalized to
the photon decay rate $\kappa$. Moreover, in the mean-field limit
the normalized amplitude $A_{out}$ of the field leaving the cavity,
as obtained by Eq.(5), simply reads
\begin{equation}
A_{out} \simeq i \sqrt T (A-Y/2).
\end{equation}
Let us now discuss separately the cases of lasing [Fig.1(b)] and CPA
[Fig.1(c)], highlighting some distinct features not considered in
Refs.\cite{StonePRL2010,StoneScience2011} and arising from
dispersive and saturation effects.\\

\par
 {\it The laser system.} The laser
configuration of Fig.1(b) corresponds to the absence of the injected
field, $Y=0$, and to an amplifying medium, $C<0$. In this case, as
is well-known laser emission occurs above threshold, for
$|C|>|C_{th}| \equiv (1+\Delta^2)/2$. The frequency
$\omega=\omega_{las}$ of the emitted radiation is obtained from the
condition $\Delta=-\theta$, which yields the well-known frequency
pulling relation
\begin{equation}
\omega_{las}=\frac{\gamma_{\perp} \omega_c+\kappa
\omega_0}{\gamma_{\perp}+\kappa}.
\end{equation}
Above threshold, the steady state solution of the normalized
intracavity power $X=|A|^2$ is given by
\begin{equation}
X=2|C|-1-\Delta^2,
\end{equation}
 and it is stable within the mean-field model. The normalized output power is simply given by
 $|A_{out}|^2=TX$.\\

 \par
{\it The CPA system.} CPA is realized for an absorbing medium
($C>0$) with an injected signal ($Y \neq 0$) of appropriate
amplitude and frequency such that the output field $A_{out}$
vanishes. Contrary to the predictions of Ref.\cite{StonePRL2010}, we
show here that the absorption-induced dispersion of the two-level
atoms breaks exact time-reversal symmetry, and the frequency at
which CPA occurs is generally different than the lasing frequency
$\omega_{las}$, given by Eq.(10). Moreover, as compared to
Ref.\cite{StonePRL2010} our model properly accounts for saturation
of the absorber, and thus can predict CPA beyond the condition of
lasing at threshold. To determine the conditions that realize a CPA,
let us first notice that the steady-state solution to Eqs.(6-8)
gives the following implicit equation between the normalized powers
$|Y|^2$ and $X=|A|^2$ of incoming and intracavity fields
\begin{equation}
|Y|^2=X \left[\left( 1+\frac{2C}{1+\Delta^2+X}\right)^2+\left(
-\theta+\frac{2C \Delta}{1+\Delta^2+X}\right)^2 \right].
\end{equation}
\begin{figure}
\includegraphics[scale=0.53]{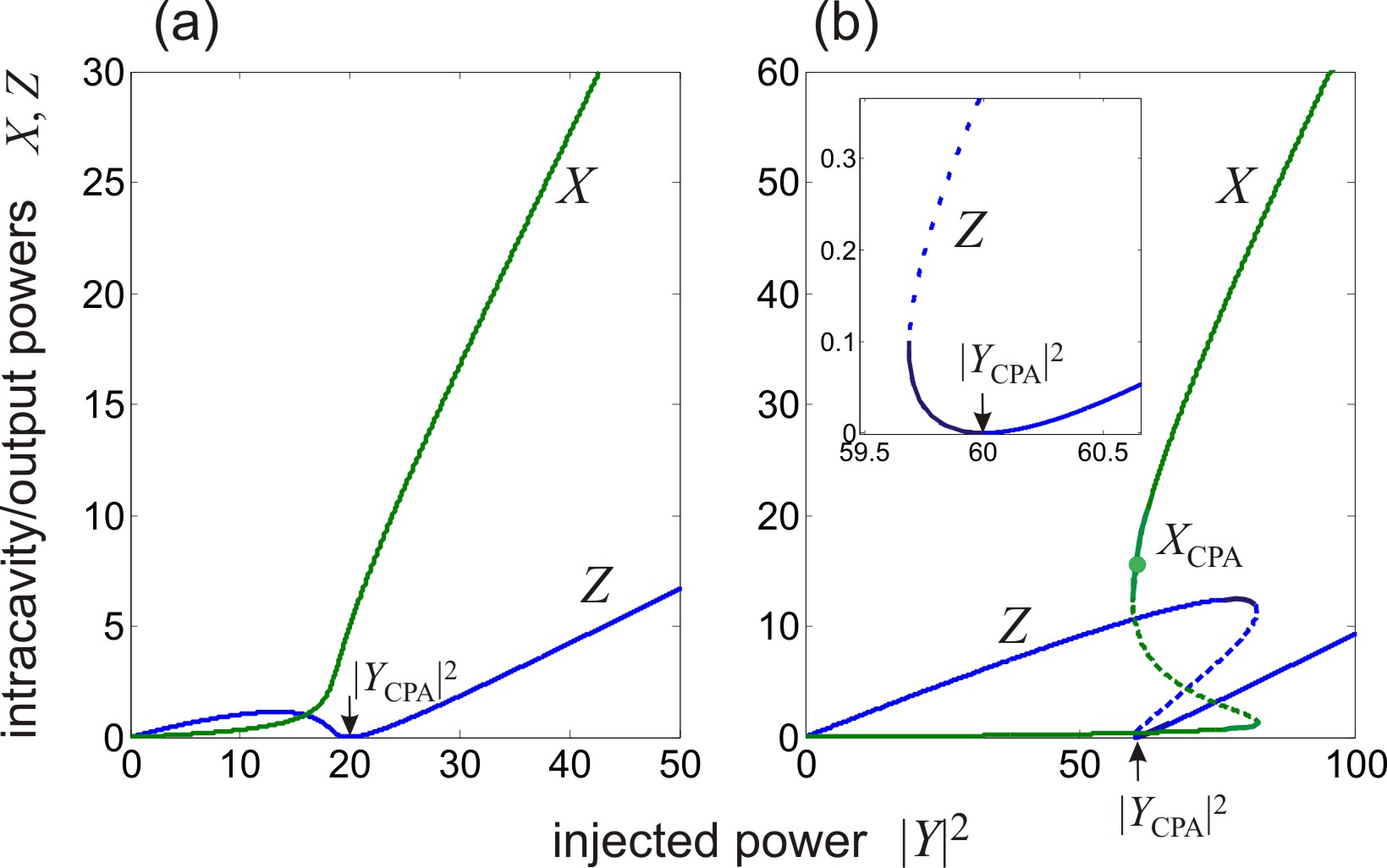}
\caption{(Color online) Behavior of the normalized intracavity power
$X=|A|^2$ and output power $Z=|A_{out}|^2/T$ versus the normalized
intensity $|Y|^2$ of the incident beam for $\Delta=\theta=0$ and for
(a) $C=3$ (monostable regime), and (b) $C=8$ (bistable regime). In
(b), the dashed curves denote the unstable branches. The value
$Y=Y_{CPA}$ corresponds to the input field amplitude that realizes
CPA. In (b) the amplitude $Y_{CPA}$ is close to the boundary of the
hysteresis cycle, as shown in the enlargement depicted in the inset
of (b).}
\end{figure}
The normalized power of the field leaving the cavity can be then
obtained using Eq.(9).  The steady-state solution (12) and its
stability have been extensively studied in the theory of optical
bistability, and the results can be briefly summarized as follows
\cite{B3}: (i) for $2C>\Delta \theta-1$ and
$(2C-\Delta\theta+1)^2(C+4\Delta \theta-4)>27C(\Delta+\theta)^2$,
the curve $X=X(|Y|^2)$ is S-shaped. Within the mean-field
approximation,  the solution is stable in the upper and lower
branches and unstable in the intermediate (negative-slope) branch.
(ii) If either one of the two previous inequalities is not
satisfied, the curve $X=X(|Y|^2)$ turns out to be single-valued and
the solution is always stable. In particular, for $C<4$ the curve is
always single-valued. As an example, in Figs.2(a) and 2(b) the
behaviors of the normalized intracavity power $X$ and output power
$Z=|A_{out}|^2/T$ versus the normalized intensity $|Y|^2$ of the
incident beam are depicted for a typical monastable [Fig.2(a)] and bistable [Fig.2(c)] cases.\\
Let us now focus our attention to the
conditions that realize a CPA. According to Eq.(9), CPA corresponds
to a steady-state solution to Eqs.(6-8) with $A=Y/2$. This leads to
the following two requirements for CPA:\\
(i) the intracavity power $X$ should take the value
$X=X_{CPA}=2C-1-\Delta^2$, which is precisely the value given by
Eq.(11); correspondingly, the normalized power $|Y|^2$ of the
injected field that realizes CPA reads
$|Y_{CPA}|^2=4X_{CPA}=4(2C-1-\Delta^2)$. In case of optical
bistability, it can be easily proven that $X_{CPA}$ is always
located onto the upper stable branch of the S-shaped curve [see, for
instance, Fig.2(b)].\\
(ii) The frequency $\omega$ of the injected
field should satisfy the condition $\theta=\Delta$.\\
Let us first discuss the former condition (i). This condition shows
that, to observe  CPA, a necessary requirement is that the
cooperation parameter $C=\alpha l /T$  must be {\it larger} than a
minimum value, which is precisely the threshold value $|C_{th}|$ of
the lasing mode once the absorber is replaced by the amplifier. If
the injected power is small enough in such a way that saturation of
absorption is negligible, i.e. $X_{CPA} \ll 1$, according to
Ref.\cite{StonePRL2010} CPA is obtained for a lossy medium
corresponding to a positive imaginary refractive index equal in
absolute value to that at the lasing threshold. For a cooperation
parameter larger than the threshold value, CPA is attained at the
intracavity power $X_{CPA}$ that makes the {\it saturated} positive
imaginary refractive index of the absorber equal in absolute value
to that at the lasing threshold. Thus, perfect absorption occurs
solely at a precise power level of the injected field, given by
$|Y|^2=|Y_{CPA}|^2=4X_{CPA}=4(2C-1-\Delta^2)$. Let us now assume
that the injected signal is adiabatically increased from $Y=0$ to
the final value $Y_{CPA}$ that realizes CPA. If the input-output
curve [Eq.(12)] is not S-shaped, CPA is obviously realized once $Y$
reaches the final value $Y_{CPA}$. This is shown, as an example, in
Fig.2(a). However, if the input-output curve [Eq.(12)] is S-shaped
and $X_{CPA}$ lies inside the bistable interval, CPA {\it is not}
attained when $Y$ reaches the value $Y_{CPA}$ because $X_{CPA}$ lies
on the upper branch of the hysteresis cycle,  as shown in Fig.2(b).
To realize CPA, one should therefore follows the bistable cycle,
first increasing the power level $|Y|^2$ above the hysteresis
threshold to switch the system from the lower to the upper stable
branch, and then decreasing the input power level down to
$|Y_{CPA}|^2$. Thus, at large values of the cooperation parameter
such that optical bistability appears, CPA could be prevented by the
appearance of the hysteresis, and its observation requires to switch
the operation into the upper branch
of the bistable curve.\\
Let us now discuss the second condition (ii), $\Delta=\theta$,
requested to observe CPA. Such a condition basically determines the
frequency $\omega=\omega_{CPA}$ of the injected field for which CPA
is observable. The main result here is that $\omega_{CPA} \neq
\omega_{las}$, where $\omega_{las}$ is given by the frequency
pulling equation (10). Such a difference is basically related to the
physical circumstance that, if the two-level amplifier is replaced
by a two-level absorber, the resonant contribution to the {\it real}
part of the refractive index changes sign as well, and the condition
$\epsilon(z) \rightarrow \epsilon^*(z)$ -invoked to explain the CPA
process as the time-reversed process of lasing at threshold
\cite{StonePRL2010}- is not strictly satisfied. In other words,
time-reversal symmetry is broken for the Maxwell-Bloch equations.
The frequency $\omega=\omega_{CPA}$ of the injected field that
yields a CPA is simply calculated by the requirement
$\Delta=\theta$. For $\kappa \neq \gamma_{\perp}$, such a condition
gives [compare with Eq.(10)]
\begin{equation}
\omega_{CPA}=\frac{\gamma_{\perp} \omega_{c}-\kappa
\omega_{0}}{\gamma_{\perp}-\kappa}.
\end{equation}
Note that, for $\gamma_{\perp} \gg \kappa$, i.e. in the good cavity
limit and for a sufficiently broadened absorption line, one has
$\omega_{CPA} \simeq \omega_{las} \simeq \omega_c$. More interesting
is the case $\gamma_{\perp}  \rightarrow \kappa$, i.e. the case
where the bandwidth of the absorption line equals the decay rate of
photons in the cold cavity. In this case the condition
$\Delta=\theta$ is never satisfied for $\omega_c \neq \omega_0$, and
it is satisfied for any frequency $\omega$ for $\omega_c=\omega_0$.
Therefore, if $\gamma_{\perp}=\kappa$ and a resonance frequency
$\omega_c$ of the cavity is exactly tuned in resonance with the
two-level atoms, the observation of CPA {\it does not require} a
precise frequency tuning of the injected field.\\

\par
{\it Conclusions.} In this Brief Report, the time-reversed process
of lasing and coherent perfect absorption proposed in recent works
\cite{StonePRL2010,StoneScience2011} have been investigated in the
framework of the semiclassical (Maxwell-Bloch) laser equations for a
homogeneously-broadened two-level medium. It has been shown that,
owing to the dispersive properties of the two-level medium, exact
time-reversal symmetry is broken and the frequency of the field at
which CPA occurs is generally different than the one of the lasing
mode. Moreover,  at a large cooperation parameter the observation of
CPA in the presence of bistability requires to operate the system in
the upper branch of the hysteresis cycle.\\

\par
This work was supported by the Italian MIUR (Grant No.
PRIN-20082YCAAK, "Analogie ottico-quantistiche in strutture
fotoniche a guida d'onda").


\begin{thebibliography}{}

\bibitem{A1}
R.H. Yan, R.J. Simes, and L.A. Coldren, IEEE Photon. Technol. Lett.
{\bf 1}, 273 (1989); K.-K. Law, R.H. Yan, J.L. Merz, and L.A.
Coldren, Appl. Phys. Lett. {\bf 56}, 1886 (1990); K.-K. Law, R.H.
Yan, L.A. Coldren, and J.L. Merz, Appl. Phys. Lett. {\bf 57}, 1345
(1990).

\bibitem{A2}
J.F. Heffernan, M.H. Moloney, J. Hegarty, J.S. Roberts, and M.
Whitehead, Appl. Phys. Lett. {\bf 58}, 2877 (1991).

\bibitem{A3}
 M.S. \"{U}nl\"{u}, K. Kishino, H. J. Liaw, and H. Morko\c{c}, J. Appl.
Phys. {\bf 71}, 4049 (1992).

\bibitem{A4}
M. Cai, O. Painter, K. J. Vahala, Phys. Rev. Lett. 85, 74 (2000); A.
Yariv, IEEE Photon. Technol. Lett. {\bf 14}, 483 (2002); J.R.
Tischler, M. S. Bradley, and V. Bulovic, Opt. Lett. {\bf 31}, 2045
(2006).

\bibitem{B1}
E. Abraham and S.D. Smith, Rep. Prog. Phys. {\bf 45}, 815 (1982).

\bibitem{B2}
H. M. Gibbs, {\it Optical Bistability: Controlling Light with Light}
(Academic, Orlando, Fla., 1985).

\bibitem{B3}
L.A. Lugiato, Prog. Opt. {\bf 21}, 69 (1984).

\bibitem{StonePRL2010}
Y. D. Chong, Li Ge, Hui Cao, and A. D. Stone, Phys. Rev. Lett. {\bf
105}, 053901 (2010); see also the viewpoint of S. Longhi, Physics
{\bf 3}, 61 (2010).

\bibitem{StoneScience2011}
W. Wan, Y. Chong, Li Ge, H. Noh, A.D. Stone, and Hui Cao, Science
{\bf 331}, 889 (2011).

\bibitem{LonghiPRA2010}
S. Longhi, Phys. Rev. A {\bf 82}, 031801(R) (2010).

\bibitem{StonePRL2011}
Y.D. Chong, Li Ge, and A.D. Stone, Phys. Rev. Lett. {\bf 106},
093902 (2011).

\bibitem{B4}
R. Bonifacio and L.A. Lugiato, Lett. Nuovo Cimento {\bf 21}, 505
(1972); R. Bonifacio and L.A. Lugiato, Phys. Rev. A {\bf 18}, 1129
(1978); M. Gronchi, V. Benza, L.A. Lugiato, P. Meystre, and M.
Sargent III, Phys. Rev. A {\bf 24}, 1419 (1981).

\bibitem{L1}
N. B. Abrahm, P. Mandel, and L.M. Narducci, Prog. Opt. {\bf 25}, 1
(1988).


\end{thebibliography}
\end{document}